\PassOptionsToPackage{unicode}{hyperref}
\PassOptionsToPackage{hyphens}{url}
\PassOptionsToPackage{dvipsnames,svgnames,x11names}{xcolor}
\documentclass[
]{article}
\usepackage{amsmath,amssymb}
\usepackage{lmodern}
\usepackage{iftex}
\ifPDFTeX
  \usepackage[T1]{fontenc}
  \usepackage[utf8]{inputenc}
  \usepackage{textcomp} 
\else 
  \usepackage{unicode-math}
  \defaultfontfeatures{Scale=MatchLowercase}
  \defaultfontfeatures[\rmfamily]{Ligatures=TeX,Scale=1}
\fi
\IfFileExists{upquote.sty}{\usepackage{upquote}}{}
\IfFileExists{microtype.sty}{
  \usepackage[]{microtype}
  \UseMicrotypeSet[protrusion]{basicmath} 
}{}
\makeatletter
\@ifundefined{KOMAClassName}{
  \IfFileExists{parskip.sty}{%
    \usepackage{parskip}
  }{
    \setlength{\parindent}{0pt}
    \setlength{\parskip}{6pt plus 2pt minus 1pt}}
}{
  \KOMAoptions{parskip=half}}
\makeatother
\usepackage{xcolor}
\usepackage{color}
\usepackage{fancyvrb}

\DefineVerbatimEnvironment{Highlighting}{Verbatim}{commandchars=\\\{\}}
\newenvironment{Shaded}{}{}

\newcommand{\BuiltInTok}[1]{\textcolor[rgb]{0.00,0.50,0.00}{#1}}

\newcommand{\CommentTok}[1]{\textcolor[rgb]{0.38,0.63,0.69}{\textit{#1}}}

\newcommand{\ConstantTok}[1]{\textcolor[rgb]{0.53,0.00,0.00}{#1}}

\newcommand{\DataTypeTok}[1]{\textcolor[rgb]{0.56,0.13,0.00}{#1}}

\newcommand{\FloatTok}[1]{\textcolor[rgb]{0.25,0.63,0.44}{#1}}
\newcommand{\FunctionTok}[1]{\textcolor[rgb]{0.02,0.16,0.49}{#1}}
\newcommand{\ImportTok}[1]{\textcolor[rgb]{0.00,0.50,0.00}{\textbf{#1}}}

\newcommand{\KeywordTok}[1]{\textcolor[rgb]{0.00,0.44,0.13}{\textbf{#1}}}
\newcommand{\NormalTok}[1]{#1}
\newcommand{\OperatorTok}[1]{\textcolor[rgb]{0.40,0.40,0.40}{#1}}

\newcommand{\PreprocessorTok}[1]{\textcolor[rgb]{0.74,0.48,0.00}{#1}}

\providecommand{\tightlist}{%
  \setlength{\itemsep}{0pt}\setlength{\parskip}{0pt}}
\newlength{\cslhangindent}
\newlength{\csllabelwidth}
\newlength{\cslentryspacingunit} 
\newenvironment{CSLReferences}[2] 
 {
  \setlength{\parindent}{0pt}
  \ifodd #1
  \let\oldpar\par
  \def\par{\hangindent=\cslhangindent\oldpar}
  \fi
  \setlength{\parskip}{#2\cslentryspacingunit}
 }%
 {}
\usepackage{calc}

\ifLuaTeX
\usepackage[bidi=basic]{babel}
\else
\usepackage[bidi=default]{babel}
\fi
\babelprovide[main,import]{american}

\def\languageshorthands#1{}
\ifLuaTeX
  \usepackage{selnolig}  
\fi
\IfFileExists{bookmark.sty}{\usepackage{bookmark}}{\usepackage{hyperref}}
\IfFileExists{xurl.sty}{\usepackage{xurl}}{} 
\hypersetup{
  pdftitle={Copulas.jl: A fully Distributions.jl-compliant copula
package},
  pdfauthor={Oskar Laverny, Santiago Jimenez},
  pdflang={en-US},
  colorlinks=true,
  linkcolor={Maroon},
  filecolor={Maroon},
  citecolor={Blue},
  urlcolor={Blue},
  pdfcreator={LaTeX via pandoc}}

\title{Copulas.jl: A fully Distributions.jl-compliant copula package}

\usepackage[affil-it]{authblk}
\usepackage{orcidlink}
\author[1%
  \ensuremath\mathparagraph]{Oskar Laverny%
    \,\orcidlink{0000-0002-7508-999X}\,%
    }
\author[2%
  ]{Santiago Jimenez%
    \,\orcidlink{0000-0002-8198-3656}\,%
    }

\affil[1]{Aix Marseille Univ, Inserm, IRD, SESSTIM, Sciences Economiques
\& Sociales de la Santé \& Traitement de l'Information Médicale, ISSPAM,
Marseille, France.}
\affil[2]{Federal University of Pernambuco}
\affil[$\mathparagraph$]{Corresponding author}
\date{16 November 2023}

\begin{document}
\maketitle

\hypertarget{summary}{%
\section{Summary}\label{summary}}

Copulas are functions that describe dependence structures of random
vectors, without describing their univariate marginals. In statistics,
the separation is sometimes useful, the quality and/or quantity of
available information on these two objects might differ. This separation
can be formally stated through Sklar's theorem:

\textbf{Theorem: existance and uniqueness of the copula
(\protect\hyperlink{ref-sklar1959fonctions}{Sklar, 1959}):} For a given
\(d\)-variate absolutely continuous random vector \(\mathbf X\) with
marginals \(X_1,...X_d\), there exists a unique function \(C\), the
copula, such that \[F(\mathbf x) = C(F_1(x_1),...,F_d(x_d)),\] where
\(F, F_1,...F_d\) are respectively the distributions functions of
\(\mathbf X, X_1,...X_d\).

Copulas are standard tools in probability and statistics, with a wide
range of applications from biostatistics, finance or medicine, to fuzzy
logic, global sensitivity and broader analysis. A few standard
theoretical references on the matter are
(\protect\hyperlink{ref-joe1997}{Joe, 1997}),
(\protect\hyperlink{ref-nelsen2006}{Nelsen, 2006}),
(\protect\hyperlink{ref-joe2014}{Joe, 2014}), and
(\protect\hyperlink{ref-durantePrinciplesCopulaTheory2015}{Durante \&
Sempi, 2015}).

The Julia package \texttt{Copulas.jl} brings most standard
copula-related features into native Julia: random number generation,
density and distribution function evaluations, fitting, construction of
multivariate models through Sklar's theorem, and many more related
functionalities. Copulas being fundamentally distributions of random
vectors, we fully comply with the
\href{https://github.com/JuliaStats/Distributions.jl}{\texttt{Distributions.jl}}
API (\protect\hyperlink{ref-djl1}{Besançon et al., 2021};
\protect\hyperlink{ref-djl2}{Lin et al., 2019}), the Julian standard for
implementation of random variables and random vectors. This complience
allows interoperability with other packages based on this API such as,
e.g., \href{https://github.com/TuringLang/Turing.jl}{\texttt{Turing.jl}}
(\protect\hyperlink{ref-turing}{Ge et al., 2018}) and several others.

\hypertarget{statement-of-need}{%
\section{Statement of need}\label{statement-of-need}}

The R package \texttt{copula}
(\protect\hyperlink{ref-r_copula_citation1}{Hofert et al., 2020};
\protect\hyperlink{ref-r_copula_citation3}{Ivan Kojadinovic \& Jun Yan,
2010}; \protect\hyperlink{ref-r_copula_citation2}{Jun Yan, 2007};
\protect\hyperlink{ref-r_copula_citation4}{Marius Hofert \& Martin
Mächler, 2011}) is the gold standard when it comes to sampling,
estimating, or simply working around dependence structures. However, in
other languages, the available tools are not as developped and/or not as
recognised. We bridge the gap in the Julian ecosystem with this
Julia-native implementation. Due to the very flexible type system in
Julia, our code expressiveness and tidyness will increase its usability
and maintenability in the long-run. Type-stability allows sampling in
arbitrary precision without requiering more code, and Julia's multiple
dispatch yields most of the below-described applications.

There are competing packages in Julia, such as
\href{https://github.com/AnderGray/BivariateCopulas.jl}{\texttt{BivariateCopulas.jl}}
which only deals with a few models in bivariate settings but has very
nice graphs, or
\href{https://github.com/iitis/DatagenCopulaBased.jl}{\texttt{DatagenCopulaBased.jl}},
which only provides sampling and does not have exactly the same models
as \texttt{Copulas.jl}. While not fully covering out both of these
package's functionality (mostly because the three projects chose
different copulas to implement), \texttt{Copulas.jl} is clearly the must
fully featured, and brings, as a key feature, the complience with the
broader ecosystem.

\hypertarget{examples}{%
\section{Examples}\label{examples}}

\hypertarget{sklardist-sampling-and-fitting-examples}{%
\subsection{\texorpdfstring{\texttt{SklarDist}: sampling and fitting
examples}{SklarDist: sampling and fitting examples}}\label{sklardist-sampling-and-fitting-examples}}

The \texttt{Distributions.jl}'s API provides a \texttt{fit} function.
You may use it to simply fit a compound model to some dataset as
follows:

\begin{Shaded}
\begin{Highlighting}[]
\ImportTok{using} \BuiltInTok{Copulas}\NormalTok{, }\BuiltInTok{Distributions}\NormalTok{, }\BuiltInTok{Random}

\CommentTok{\# Define the marginals and the copula, then use Sklar\textquotesingle{}s theorem:}
\NormalTok{X1 }\OperatorTok{=} \FunctionTok{Gamma}\NormalTok{(}\FloatTok{2}\NormalTok{,}\FloatTok{3}\NormalTok{)}
\NormalTok{X2 }\OperatorTok{=} \FunctionTok{Pareto}\NormalTok{(}\FloatTok{0.5}\NormalTok{)}
\NormalTok{X3 }\OperatorTok{=} \FunctionTok{Binomial}\NormalTok{(}\FloatTok{10}\NormalTok{,}\FloatTok{0.8}\NormalTok{)}
\NormalTok{C }\OperatorTok{=} \FunctionTok{ClaytonCopula}\NormalTok{(}\FloatTok{3}\NormalTok{,}\FloatTok{0.7}\NormalTok{)}
\NormalTok{X }\OperatorTok{=} \FunctionTok{SklarDist}\NormalTok{(C,(X1,X2,X3))}

\CommentTok{\# Sample from the model: }
\NormalTok{x }\OperatorTok{=} \FunctionTok{rand}\NormalTok{(D,}\FloatTok{1000}\NormalTok{)}

\CommentTok{\# You may estimate the model as follows: }
\NormalTok{Dhat }\OperatorTok{=} \FunctionTok{fit}\NormalTok{(SklarDist\{FrankCopula,}\DataTypeTok{Tuple}\NormalTok{\{Gamma,Normal,Binomial\}\}, x)}
\CommentTok{\# Although you\textquotesingle{}ll probbaly get a bad fit !}
\end{Highlighting}
\end{Shaded}

The API does not fix the fitting procedure, and only loosely specify it,
thus the implemented default might vary on the copula. If you want more
control, you may turn to bayesian estimation using \texttt{Turing.jl}
(\protect\hyperlink{ref-turing}{Ge et al., 2018}):

\begin{Shaded}
\begin{Highlighting}[]
\ImportTok{using} \BuiltInTok{Turing}
\PreprocessorTok{@model} \KeywordTok{function} \FunctionTok{model}\NormalTok{(dataset)}
  \CommentTok{\# Priors}
\NormalTok{  theta }\OperatorTok{\textasciitilde{}} \FunctionTok{TruncatedNormal}\NormalTok{(}\FloatTok{1.0}\NormalTok{, }\FloatTok{1.0}\NormalTok{, }\FloatTok{theta}\NormalTok{, }\ConstantTok{Inf}\NormalTok{)}
\NormalTok{  gamma }\OperatorTok{\textasciitilde{}} \FunctionTok{TruncatedNormal}\NormalTok{(}\FloatTok{1.0}\NormalTok{, }\FloatTok{1.0}\NormalTok{, }\FloatTok{0.25}\NormalTok{, }\ConstantTok{Inf}\NormalTok{)}
\NormalTok{  eta }\OperatorTok{\textasciitilde{}} \FunctionTok{Beta}\NormalTok{(}\FloatTok{1}\NormalTok{,}\FloatTok{1}\NormalTok{)}
\NormalTok{  delta }\OperatorTok{\textasciitilde{}} \FunctionTok{Exponential}\NormalTok{(}\FloatTok{1}\NormalTok{)}

  \CommentTok{\# Define the model through Sklar\textquotesingle{}s theorem: }
\NormalTok{  X1 }\OperatorTok{=} \FunctionTok{Gamma}\NormalTok{(}\FloatTok{2}\NormalTok{,theta)}
\NormalTok{  X2 }\OperatorTok{=} \FunctionTok{Pareto}\NormalTok{(gamma)}
\NormalTok{  X3 }\OperatorTok{=} \FunctionTok{Binomial}\NormalTok{(}\FloatTok{10}\NormalTok{,eta)}
\NormalTok{  C }\OperatorTok{=} \FunctionTok{ClaytonCopula}\NormalTok{(}\FloatTok{3}\NormalTok{,delta)}
\NormalTok{  X }\OperatorTok{=} \FunctionTok{SklarDist}\NormalTok{(C,(X1,X2,X3))}

  \CommentTok{\# Add the loglikelyhood to the model : }
\NormalTok{  Turing.Turing.}\PreprocessorTok{@addlogprob}\NormalTok{! }\FunctionTok{loglikelihood}\NormalTok{(D, dataset)}
\KeywordTok{end}
\end{Highlighting}
\end{Shaded}

\hypertarget{the-archimedean-api}{%
\subsection{The Archimedean API}\label{the-archimedean-api}}

Archimedean copulas are a huge family of copulas that has seen a lot of
theoretical work. Among others, you may take a look at
(\protect\hyperlink{ref-mcneilMultivariateArchimedeanCopulas2009b}{McNeil
\& Nešlehová, 2009}). We use
\href{https://github.com/lrnv/WilliamsonTransforms.jl/}{\texttt{WilliamsonTransformations.jl}}'s
implementation of the Williamson \(d\)-transfrom to sample from any
archimedean copula, including for example the \texttt{ClaytonCopula}
with negative dependence parameter in any dimension, which is a first to
our knowledge.

The API is consisting of the folloiwng functions:

\begin{Shaded}
\begin{Highlighting}[]
\FunctionTok{phi}\NormalTok{(C}\OperatorTok{::}\DataTypeTok{MyArchimedean}\NormalTok{, t) }\CommentTok{\# Generator}
\FunctionTok{williamson\_dist}\NormalTok{(C}\OperatorTok{::}\DataTypeTok{MyArchimedean}\NormalTok{) }\CommentTok{\# Williamson d{-}transform}
\end{Highlighting}
\end{Shaded}

So that implementing your own archimedean copula only requires to subset
the \texttt{ArchimedeanCopula} type and provide your generator as
follows:

\begin{Shaded}
\begin{Highlighting}[]
\KeywordTok{struct}\NormalTok{ MyUnknownArchimedean\{d,T\} }\OperatorTok{\textless{}:}\DataTypeTok{ ArchimedeanCopula\{d\}}
\NormalTok{    theta}\OperatorTok{::}\DataTypeTok{T}
\KeywordTok{end}
\FunctionTok{phi}\NormalTok{(C}\OperatorTok{::}\DataTypeTok{MyUnknownArchimedean}\NormalTok{,t) }\OperatorTok{=} \FunctionTok{exp}\NormalTok{(}\OperatorTok{{-}}\NormalTok{t}\OperatorTok{*}\NormalTok{C.theta)}
\end{Highlighting}
\end{Shaded}

The obtained model can be used as follows:

\begin{Shaded}
\begin{Highlighting}[]
\NormalTok{C }\OperatorTok{=} \FunctionTok{MyUnknownCopula}\DataTypeTok{\{2,Float64\}}\NormalTok{(}\FloatTok{3.0}\NormalTok{)}
\NormalTok{spl }\OperatorTok{=} \FunctionTok{rand}\NormalTok{(C,}\FloatTok{1000}\NormalTok{)   }\CommentTok{\# sampling}
\FunctionTok{cdf}\NormalTok{(C,spl)           }\CommentTok{\# cdf}
\FunctionTok{pdf}\NormalTok{(C,spl)           }\CommentTok{\# pdf}
\FunctionTok{loglikelihood}\NormalTok{(C,spl) }\CommentTok{\# llh}
\end{Highlighting}
\end{Shaded}

The following functions have defaults but can be overridden for
performance:

\begin{Shaded}
\begin{Highlighting}[]
\FunctionTok{phi_inv}\NormalTok{(C}\OperatorTok{::}\DataTypeTok{MyArchimedean}\NormalTok{, t) }\CommentTok{\# Inverse of phi}
\FunctionTok{phi_1}\NormalTok{(C}\OperatorTok{::}\DataTypeTok{MyArchimedean}\NormalTok{, t) }\CommentTok{\# first defrivative of phi}
\FunctionTok{phi_d}\NormalTok{(C}\OperatorTok{::}\DataTypeTok{MyArchimedean}\NormalTok{,t) }\CommentTok{\# dth defrivative of phi}
\FunctionTok{tau}\NormalTok{(C}\OperatorTok{::}\DataTypeTok{MyArchimedean}\NormalTok{) }\CommentTok{\# Kendall tau}
\FunctionTok{tau_inv}\NormalTok{(}\OperatorTok{::}\DataTypeTok{Type\{MyArchimedean\}}\NormalTok{,tau) }\OperatorTok{=} \CommentTok{\# Inverse kendall tau}
\FunctionTok{fit}\NormalTok{(}\OperatorTok{::}\DataTypeTok{Type\{MyArchimedean\}}\NormalTok{,data) }\CommentTok{\# fitting.}
\end{Highlighting}
\end{Shaded}

\hypertarget{broader-ecosystem}{%
\subsection{Broader ecosystem}\label{broader-ecosystem}}

The package is starting to get used in several other places of the
ecosystem. Among others, we noted:

\begin{itemize}
\tightlist
\item
  The package
  \href{https://github.com/SciML/GlobalSensitivity.jl}{\texttt{GlobalSensitivity.jl}}
  exploit \texttt{Copulas.jl} to provide Shapley effects implementation,
  see
  \href{https://docs.sciml.ai/GlobalSensitivity/stable/tutorials/shapley/}{this
  documentation}.
\item
  \href{https://github.com/JuliaActuary/EconomicScenarioGenerators.jl}{\texttt{EconomicScenarioGenerators.jl}}
  uses depndence structures between financial assets.
\end{itemize}

\hypertarget{acknowledgments}{%
\section{Acknowledgments}\label{acknowledgments}}

\hypertarget{references}{%
\section*{References}\label{references}}
\addcontentsline{toc}{section}{References}

\hypertarget{refs}{}
\begin{CSLReferences}{1}{0}
\leavevmode\vadjust pre{\hypertarget{ref-djl1}{}}%
Besançon, M., Papamarkou, T., Anthoff, D., Arslan, A., Byrne, S., Lin,
D., \& Pearson, J. (2021). Distributions.jl: Definition and modeling of
probability distributions in the JuliaStats ecosystem. \emph{Journal of
Statistical Software}, \emph{98}(16), 1--30.
\url{https://doi.org/10.18637/jss.v098.i16}

\leavevmode\vadjust pre{\hypertarget{ref-durantePrinciplesCopulaTheory2015}{}}%
Durante, F., \& Sempi, C. (2015). \emph{Principles of copula theory}.
{Chapman and Hall/CRC}. \url{https://doi.org/10.1201/b18674}

\leavevmode\vadjust pre{\hypertarget{ref-turing}{}}%
Ge, H., Xu, K., \& Ghahramani, Z. (2018). Turing: A language for
flexible probabilistic inference. \emph{International Conference on
Artificial Intelligence and Statistics, {AISTATS} 2018, 9-11 April 2018,
Playa Blanca, Lanzarote, Canary Islands, Spain}, 1682--1690.
\url{http://proceedings.mlr.press/v84/ge18b.html}

\leavevmode\vadjust pre{\hypertarget{ref-r_copula_citation1}{}}%
Hofert, M., Kojadinovic, I., Maechler, M., \& Yan, J. (2020).
\emph{Copula: Multivariate dependence with copulas}.
\url{https://CRAN.R-project.org/package=copula}

\leavevmode\vadjust pre{\hypertarget{ref-r_copula_citation3}{}}%
Ivan Kojadinovic, \& Jun Yan. (2010). Modeling multivariate
distributions with continuous margins using the {copula} {R} package.
\emph{Journal of Statistical Software}, \emph{34}(9), 1--20.
\url{https://doi.org/10.18637/jss.v034.i09}

\leavevmode\vadjust pre{\hypertarget{ref-joe1997}{}}%
Joe, H. (1997). \emph{Multivariate models and multivariate dependence
concepts}. {CRC press}. \url{https://doi.org/10.1201/9780367803896}

\leavevmode\vadjust pre{\hypertarget{ref-joe2014}{}}%
Joe, H. (2014). \emph{Dependence modeling with copulas}. {CRC press}.

\leavevmode\vadjust pre{\hypertarget{ref-r_copula_citation2}{}}%
Jun Yan. (2007). Enjoy the joy of copulas: With a package {copula}.
\emph{Journal of Statistical Software}, \emph{21}(4), 1--21.
\url{https://doi.org/10.18637/jss.v021.i04}

\leavevmode\vadjust pre{\hypertarget{ref-djl2}{}}%
Lin, D., White, J. M., Byrne, S., Bates, D., Noack, A., Pearson, J.,
Arslan, A., Squire, K., Anthoff, D., Papamarkou, T., Besançon, M.,
Drugowitsch, J., Schauer, M., \& contributors, other. (2019).
\emph{{JuliaStats/Distributions.jl: a Julia package for probability
distributions and associated functions}}.
\url{https://doi.org/10.5281/zenodo.2647458}

\leavevmode\vadjust pre{\hypertarget{ref-r_copula_citation4}{}}%
Marius Hofert, \& Martin Mächler. (2011). Nested archimedean copulas
meet {R}: The {nacopula} package. \emph{Journal of Statistical
Software}, \emph{39}(9), 1--20.
\url{https://doi.org/10.18637/jss.v039.i09}

\leavevmode\vadjust pre{\hypertarget{ref-mcneilMultivariateArchimedeanCopulas2009b}{}}%
McNeil, A. J., \& Nešlehová, J. (2009). Multivariate {Archimedean}
copulas, d -monotone functions and L1 -norm symmetric distributions.
\emph{The Annals of Statistics}, \emph{37}(5B), 3059--3097.
\url{https://doi.org/10.1214/07-AOS556}

\leavevmode\vadjust pre{\hypertarget{ref-nelsen2006}{}}%
Nelsen, R. B. (2006). \emph{An introduction to copulas} (2nd ed).
{Springer}. ISBN:~978-0-387-28659-4

\leavevmode\vadjust pre{\hypertarget{ref-sklar1959fonctions}{}}%
Sklar, A. (1959). Fonction de r{é}partition dont les marges sont
donn{é}es. \emph{Inst. Stat. Univ. Paris}, \emph{8}, 229--231.

\end{CSLReferences}

\end{document}